\documentclass[12pt,letterpaper]{article}
 
\usepackage{amsmath}
\allowdisplaybreaks[1]

\usepackage{graphicx}
\usepackage{subfig}

\usepackage{amssymb}

\usepackage{longtable}

\def\be{\begin{equation}}
\def\ee{\end{equation}}
\def\ba{\begin{align}}
\def\ea{\end{align}}

\newcommand{\eq}[1]{(\ref{#1})}

\newcommand{\ket}[1]{\left|#1\right>}
\newcommand{\bra}[1]{\left<#1\right|}

\newcommand{\comment}[1]{}

\begin{document}


\begin{titlepage}
~
\vskip 1in
\begin{center}
{\Large
{SFT Action for Separated D-branes}}
\vskip 0.5in
{Matheson Longton}
\vskip 0.3in
{\it 
Department of Physics and Astronomy\\
University of British Columbia
Vancouver, Canada}
\end{center}

\vskip 0.5in
\begin{abstract}
We present an action for Cubic String Field Theory with one embedding 
coordinate treated separately.  We truncate the action at level 
(3,9), but unlike many other works we do not impose twist symmetry.  
We also allow arbitrary zero-modes for the direction considered 
special.  Our action provides a starting point for the study of 
numerous configurations of D-branes.
\end{abstract}
\end{titlepage}




In many studies of bosonic String Field Theory, we single out one 
direction, which we will call $X^{25}$,
as special and assume a rotational symmetry in the other 25.  
Solutions involving translations, lumps, or rolling tachyons will all 
satisfy these assumptions, as will the simpler case of D-brane decay.  The 
action presented here was derived for \cite{Karczmarek:2012pn} and 
complements that work, but is also 
valid for other situations.  We can use it to study marginal 
deformations, or solutions with non-zero momentum in the $X^{25}$ 
direction such as lumps.  Because we have not imposed twist symmetry, 
it is possible to examine the spectrum of states around a solution, 
including states which are not necessary while finding the solution.  
Solutions based on twist-odd 
operators could also be studied.  While we stop at the relatively low 
level (3,9), this is still high enough for a reasonable approximation 
and physical interpretation of most systems.

In level-truncated studies of string field theory, the string field 
is expanded in a basis of conformal primaries and their descendants, 
and then truncated at some eigenvalue of the virasoro zero-mode 
operator.  Our intended 
theory is a collection of D24-branes whose transverse directions are all 
aligned, but which are not necessarily coincident.  Singling out the 
$X^{25}$ direction is the main assumption we have made on our string 
field.  Any system of bosonic D-branes with a rotational symmetry in 
25 of the 26 dimensions should be described at level 3 by this 
potential.  The string field we use will be
\begin{multline} \label{eq.appendix.stringfield}
\ket{\Phi}=\sum_{d}\left(t_{d}c_{1}+h_{d}c_{0}+u_{d}c_{-1}+v_{d}L^{\prime}_{-2}c_{1}+w_{d}L^{25}_{-2}c_{1}+o_{d}(b_{-2}c_{-1}c_{1}-2c_{-2})\right.\\
+\left.\tilde{o}_{d}(b_{-2}c_{-1}c_{1}+2c_{-2})+p_{d}L_{-3}^{\prime}c_{1}+q_{d}L_{-3}^{25}c_{1}+\ldots\right)\ket{0;d}\\
+\sum_{d}\left(x_{d}c_{1}+f_{d}L_{-1}^{25}c_{1}+r_{d}c_{-1}+s_{d}L^{\prime}_{-2}c_{1}+y_{d}L^{25}_{-2}c_{1}+z_{d}L^{25}_{-1}L^{25}_{-1}c_{1}+\ldots\right)\alpha^{25}_{-1}\ket{0;d}
\end{multline}
where $L_{n}^{\prime}=\sum_{\mu=0}^{24}L_{n}^{\mu}$ contains the sum over the 
other 25 directions of matter oscillators.  The ghost CFT is handled 
explicitly in terms of $b_{n}$ and $c_{n}$ operators.  We will often 
label the states by their associated coefficients, such as $w$ 
referring to the state $L_{-2}^{25}c_{1}\ket{0;d}$.  The vacuum 
$\ket{0;d}$ is defined to have the eigenvalue 
$\alpha_{0}^{25}\ket{0;d}=d\ket{0;d}$, and
$\alpha_{0}^{\mu\neq 25}\ket{0;d}=0$.  This notation for the vacua is 
not what would typically be used once a target solution has been 
chosen, as there may be distinct vacuum states with the same 
$\alpha_{0}^{25}$ eigenvalue.  The case of non-trivial 
zero-mode eigenvalue in the rotationally symmetric directions has not 
been considered.

For the intended situation of multiple separated D-branes \cite{Karczmarek:2012pn}, 
the eigenvalues $d$ are zero for strings which begin and end on the 
same brane, but are proportional to the brane separation for stretched 
strings.  The sum over vacuum states now becomes a sum over 
Chan-Paton indices.
\begin{multline} \label{eq.appendix.cpstringfield}
\ket{\Phi}=\sum_{i,j}\left(t_{ij}c_{1}+h_{ij}c_{0}+u_{ij}c_{-1}+v_{ij}L^{\prime}_{-2}c_{1}+w_{ij}L^{25}_{-2}c_{1}+o_{ij}(b_{-2}c_{-1}c_{1}-2c_{-2})\right.\\
+\left.\tilde{o}_{ij}(b_{-2}c_{-1}c_{1}+2c_{-2})+p_{ij}L_{-3}^{\prime}c_{1}+q_{ij}L_{-3}^{25}c_{1}+\ldots\right)\ket{ij}\\
+\sum_{i,j}\left(x_{ij}c_{1}+f_{ij}L_{-1}^{25}c_{1}+r_{ij}c_{-1}+s_{ij}L^{\prime}_{-2}c_{1}+y_{ij}L^{25}_{-2}c_{1}+z_{ij}L^{25}_{-1}L^{25}_{-1}c_{1}+\ldots\right)\alpha^{25}_{-1}\ket{ij}
\end{multline}
Here $\ket{ij}=\ket{0;d_{i}-d_{j}}$ represents the vacuum state of a 
string which begins on brane $i$ and ends on brane $j$.  The 
parameters $d_{i}$ are related to the physical coordinates of the 
D-branes in the transverse direction, $\hat{d_{i}}$, by 
$d_{i}=\frac{\hat{d_{i}}}{\sqrt{2\alpha^{\prime}}\pi}$.  Once we have chosen an 
explicit form of the string field we can also determine which states 
are twist even or odd, and impose a condition on the string field to 
guarantee that the action remains real.  In general the reality 
condition is $\Phi=bpz^{-1}(\Phi^{\dagger})$ \cite{Gaberdiel:1997ia}, 
but for our system it can be shown that this is related to twist 
\cite{Karczmarek:2012pn}.  In fact, all twist-odd operators have 
hermitian matrix coefficients, while all twist-even operators have 
anti-hermitian ones.  We must be careful to distinguish between the 
twist eigenvalue 
of the operator and the state, because the vacuum no longer 
transforms trivially under twist.  To find the twist behaviour of a 
state, we must first decompose it into symmetric and anti-symmetric 
parts, as they will have opposite twist eigenvalues.  The symmetric 
vacuum $\ket{ij}+\ket{ji}$ is twist-odd, as is typical for the string 
field theory vacuum, while $\ket{ij}-\ket{ji}$ is now twist-even.
Further properties of the string field, including a symmetry specific 
to two D-branes and an alternate choice of basis, have also been 
explored in \cite{Karczmarek:2012pn}.

With the string field defined, we can now move on to examine the 
action.  The quadratic term is a function of the zero-mode 
eigenvalue $d$, and for 
every set of three fields there is a cubic 
term which is a function of the three zero-modes of the strings.  For 
the study of separated D-branes or non-zero momentum states these 
zero-modes must sum to 0, but we will leave them unconstrained.
The action can be written as 
\be \label{eq.appendix.act-defn}
S=-\frac{1}{2}\sum_{l,m}A_{lm}(d)\phi^{(l)\dagger}_{d}\phi^{(m)}_{d}
-\sum_{l,m,n}B_{lmn}(d_{1},d_{2},d_{3})\left(\frac{4}{3\sqrt{3}}\right)^{\frac{1}{2}(d_{1}^{2}+d_{2}^{2}+d_{3}^{2})}\phi^{(l)}_{d_{1}}\phi^{(m)}_{d_{2}}\phi^{(n)}_{d_{3}},
\comment{V=\pi^{2}\sum_{l,m}A_{lm}(d_{j}-d_{i})\phi^{(l)}_{ij}\phi^{(m)}_{ji}
+2\pi^{2}\sum_{l,m,n}B_{lmn}(d_{1},d_{2},d_{3})\left(\frac{4}{3\sqrt{3}}\right)^{\frac{1}{2}(d_{1}^{2}+d_{2}^{2}+d_{3}^{2})}\phi^{(l)}_{ij}\phi^{(m)}_{jk}\phi^{(n)}_{ki},}
\ee
where $l$, $m$, and $n$ run over the set of all fields we are 
considering.  The purpose of this 
note is then to explicitly list the coefficient functions $A_{lm}$ 
and $B_{lmn}$ for the first few levels.  For our special case 
of separated D-branes, where the vacuum states are represented by 
Chan-Paton factors, the cubic terms use 
$d_{1}=d_{i}-d_{j},\,d_{2}=d_{j}-d_{k},\,d_{3}=d_{k}-d_{i}$, and the 
field associated with such a state has the appropriate indices 
$\phi^{(t)}_{d_{1}}=t_{ij}$.  For example, the 
action with only the fields $t_{ij}$ and $x_{ij}$ would contain the 
terms 
\begin{equation*}
S=-\frac{1}{2}\left(A_{tt}t_{ij}t_{ji}+A_{xx}x_{ij}x_{ji}\right) 
-\left(B_{ttt}t_{ij}t_{jk}t_{ki}+3B_{ttx}t_{ij}t_{jk}x_{ki}+3B_{txx}t_{ij}x_{jk}x_{ki}+B_{xxx}x_{ij}x_{jk}x_{ki}\right)
\end{equation*}
where the separation parameters of the coefficients were omitted for 
brevity.  
The cubic term of the action is usually written with a factor of $\frac{1}{3}$ in 
front of it, but we have defined the coefficients $B_{lmn}$ to 
include that so that \eq{eq.appendix.act-defn} is correct as written.  
For studies of tachyon condensation, a more useful quantity than the 
action is the 
potential rescaled to units where the mass of the brane is 1.  This is 
given by $V=-2\pi^{2}S$.  The energy of the true vacuum solution will 
appear as $-1$ in these units.

Cubic terms involving distinct fields such as $xut$ can all be related to a 
term with the operators in a fixed order such as $txu$.
While there are in principle as many as six orderings of each set of 
fields, it can 
be shown using the twist operator that we only need to calculate one of 
them \cite{Karczmarek:2012pn}.  Cyclic permutations of the fields do 
not change the coefficients $B_{lmn}$ except to appropriately relabel 
the parameters: 
$B_{lmn}(d_{1},d_{2},d_{3})=B_{mnl}(d_{2},d_{3},d_{1})=B_{nlm}(d_{3},d_{1},d_{2})$.  This 
results in a combinatoric factor of 3 whenever two of the fields are 
the same.  When all three fields are different, the other distinct 
ordering is related to the first by twist.  Letting $\Omega_{l}$ be 
the twist eigenvalue of the operator $\phi^{(l)}$, we can write 
that 
\be
B_{lmn}(d_{1},d_{2},d_{3})=-\Omega_{l}\Omega_{m}\Omega_{n}B_{nml}(-d_{3},-d_{2},-d_{1}).
\ee
For the 
case of the separated D-branes the result of adding all 6 orderings 
of distinct fields which satisfy the reality condition is
\be
6B_{lmn}(d_{1},d_{2},d_{3})\mathrm{Re}(\phi^{(l)}_{ij}\phi^{(m)}_{jk}\phi^{(n)}_{ki}).
\ee

Similarly,
we know from the properties of the inner product and the hermicity of 
$Q_{B}$ that $A_{lm}=\pm A_{ml}$.  The quadratic coefficients are 
defined by 
$\bra{\phi^{(l)};d}\phi^{(l)\dagger}_{d}Q_{B}\phi^{(m)}_{d}\ket{\phi^{(m)};d}=A_{lm}(d)\phi^{(l)\dagger}_{d}\phi^{(m)}_{d}$, 
so we know from the basic properties of the inner product and the reality of 
the action that the sign $A_{lm}$ picks up on exchanging its indices 
will be the one that makes the whole kinetic term the same for both 
orderings of the operators.  The quadratic coefficients in table 
\ref{tab.appendix.quadratic} have been given with both orderings for 
clarity.  

The action used for the calculation of lump solutions in 
\cite{Moeller:2000jy} can be reproduced using our coefficients.  The 
main difference to take into account is the choice of vacuum.  Since 
that system has a $p\rightarrow -p$ symmetry, the authors chose the 
vacuum 
$\frac{1}{2}\left(\ket{0;\frac{\sqrt{2}n}{R}}+\ket{0;\frac{-\sqrt{2}n}{R}}\right)$ 
with momentum $|p|=\frac{n}{R}$.  For every field with non-zero 
momentum, we will therefore pick up a factor of $\frac{1}{2}$, in 
addition to an overall combinatoric factor of $2$ from the 
two ways to match up momentum eigenstates in order to get a total 
momentum of $0$.  If we use our field definitions we can now simply 
plug in the appropriate values for $d_{1}$, $d_{2}$, and $d_{3}$ to 
the coefficient functions $B_{lmn}$ and get the action.  If we want 
to use the fields defined in \cite{Moeller:2000jy}, however, we must 
adjust for a few differences in naming the fields seen in table 
\ref{tab.appendix.fieldcompare} and also 
note that their $z$ term differs from our $f$ term by a factor of 
$\alpha_{0}$.  For the cubic part of the action we can multiply 
by the appropriate $d$ for each $z$ field appearing.  The quadratic 
term has an additional complication due to the fact that $\alpha_{0}$ 
is $bpz$ odd.  In addition to multiplying by a factor of $d$ for each 
$z$ field, we must include an extra sign every time $z$ appears on 
the left, as the index $l$ rather than $m$.  The action of 
\cite{Moeller:2000jy} can then be reproduced up to level 3 using our 
coefficients.

\begin{table} \begin{center}
\begin{tabular}{|r|ccccccccccccccc|}
	\hline
	Here&$t$&$x$&$h$&$u$&$v$&$w$&$f$&$o$&$\tilde{o}$&$p$&$q$&$r$&$s$&$y$&$z$ \\ \hline
	\cite{Sen:2000hx}&$t$&$a_{s}$&NA&$u$&$w$&$v$&NA&NA&NA&NA&NA&$s$&$r$&$\bar{r}$&$y$ \\ \hline
	\cite{Moeller:2000jy}&$t$&NA&NA&$u$&$w$&$v$&$z^{\ast}$&NA&NA&NA&NA&NA&NA&NA&NA \\
	\hline
\end{tabular}
\caption{A comparison of field definitions with two other works.\\
	{\footnotesize *There is a $d$-dependent difference in normalization for this field.}}\label{tab.appendix.fieldcompare}
\end{center}
\end{table}

To clarify this relationship, we will give an example.  We 
can compute the coupling of $twf$ using our action 
and compare it to the coupling of $tvz$ in \cite{Moeller:2000jy}.  To 
make this more 
precise, we must choose which momentum to assign to each field.  
$t_{0}w_{1}f_{1}$ is a good choice, and immediately gives us 
\be
B_{twf}\left(0,\frac{\sqrt{2}}{R},-\frac{\sqrt{2}}{R}\right)
=\frac{\sqrt{6}}{27R^{5}}-\frac{209\sqrt{6}}{432R^{3}}-\frac{2359\sqrt{6}}{3456R}.
\ee
As described above, we must now multiply by $\frac{2d_{3}}{4}$ in 
order to deal with the combinatorics and the difference in field 
definitions, as well as the factor of 
$\left(\frac{4}{3\sqrt{3}}\right)^{\frac{2}{R^{2}}}$ which multiplies 
the coefficient $B_{lmn}$ in 
\eq{eq.appendix.act-defn}.  Finally, the fields are distinct, so we 
need a combinatoric factor of 6.  We find that the coupling of 
$t_{0}v_{1}z_{1}$ in \cite{Moeller:2000jy} should be 
\be
\left(-\frac{\sqrt{3}}{R^{6}}+\frac{209\sqrt{3}}{432R^{4}}+\frac{2359\sqrt{3}}{3456R^{2}}\right)\left(\frac{4}{3\sqrt{3}}\right)^{\frac{2}{R^{2}}}
\ee
which is indeed what was used.  An example of a quadratic coupling 
would be $fw$, or $z_{1}v_{1}$ in terms of the other field 
definitions.  We see from table \ref{tab.appendix.quadratic} that 
\be
A_{fw}\left(\frac{\sqrt{2}}{R}\right)=-\frac{3}{2}\frac{\sqrt{2}}{R}\left(\frac{2}{R^{2}}+2\right)
=-3\sqrt{2}\left(\frac{1}{R^{3}}+\frac{1}{R}\right).
\ee
We must now multiply by $\frac{2}{8}$ for the two fields 
and the two ways to conserve momentum with these two fields, as well 
as the factor multiplying $A_{lm}$ in \eq{eq.appendix.act-defn}.  There is 
then a factor of $-\alpha_{0}=-\frac{\sqrt{2}}{R}$ from the change of 
field definition for $bpz(f)$, giving us the result 
\be
\frac{3}{2}\left(\frac{1}{R^{4}}+\frac{1}{R^{2}}\right).
\ee
When we look at the coefficient in \cite{Moeller:2000jy} we see 
$\frac{3}{R^{2}}+\frac{3}{R^{4}}$, which includes both this and the 
identical $v_{1}z_{1}$ term.  

The action without the matter zero modes, equivalent to $A_{lm}(0)$ 
and $B_{lmn}(0,0,0)$, was given in 
\cite{Sen:2000hx} and the action for lump solutions, including terms 
with non-zero momentum, was given in \cite{Moeller:2000jy}, so the 
brief dictionary of fields in table 
\ref{tab.appendix.fieldcompare} includes the fields used here as well 
as in both those works.  Our coefficients are more general, 
as we include arbitrary zero-modes instead of assuming that they 
vanish or are for specific momentum states.  We also include the 
states which were neglected by these works because they were twist 
odd.  The quadratic coefficients 
$A_{lm}(d)$ are presented in table \ref{tab.appendix.quadratic}, and 
the  rest of this note consists of the cubic terms 
$B_{lmn}(d_{1},d_{2},d_{3})$ up to level (2,6) because the higher 
level terms we have calculated would fill 30 pages more and are not 
very enlightening to view.  For a form of both 
the quadratic and cubic terms suitable for use by computer algebra 
systems, download the ancillary file ``osft-couplings.txt'' available 
with this note. This also includes the higher level terms up to level 
(3,9). 

\begin{table}[ht]\begin{center}\footnotesize\begin{tabular}{@{}|c@{}c|l|}
\hline
$l$ & $m$ & $A_{lm}(d)$ \\ \hline

$t$ & $t$ & $1/2\,{d}^{2}-1
$ \\ 
$x$ & $x$ & $1/2\,{d}^{2}
$ \\ 
$h$ & $h$ & $-2
$ \\ 
$u$ & $u$ & $-1-1/2\,{d}^{2}
$ \\ 
$v$ & $v$ & ${\frac {25}{2}}+{\frac {25}{4}}\,{d}^{2}
$ \\ 
$w$ & $w$ & $1/4\, ( 4\,{d}^{2}+1 )  ( {d}^{2}+2 ) 
$ \\ 
$w$ & $f$ & $3/2\,d ( {d}^{2}+2 ) 
$ \\ 
$f$ & $w$ & $-3/2\,d ( {d}^{2}+2 ) 
$ \\ 
$f$ & $f$ & $- ( {d}^{2}+2 )  ( {d}^{2}+1 ) 
$ \\ 
$o$ & $o$ & $8+2\,{d}^{2}
$ \\ 
$\tilde{o}$ & $\tilde{o}$ & $-8-2\,{d}^{2}
$ \\ 
$p$ & $p$ & $-100-25\,{d}^{2}
$ \\ 
$q$ & $q$ & $-1/2\, ( 3\,{d}^{2}+2 )  ( 4+{d}^{2} ) 
$ \\ 
$q$ & $y$ & $-5/2\,d ( 4+{d}^{2} ) 
$ \\ 
$y$ & $q$ & $5/2\,d ( 4+{d}^{2} ) 
$ \\ 
$q$ & $z$ & $-6\,d ( 4+{d}^{2} ) 
$ \\ 
$z$ & $q$ & $6\,d ( 4+{d}^{2} ) 
$ \\ 
$r$ & $r$ & $-2-1/2\,{d}^{2}
$ \\ 
$s$ & $s$ & ${\frac {25}{4}}\,{d}^{2}+25
$ \\ 
$y$ & $y$ & $1/4\, ( 4+{d}^{2} )  ( 4\,{d}^{2}+9 ) 
$ \\ 
$y$ & $z$ & $3/2\, ( 3\,{d}^{2}+2 )  ( 4+{d}^{2} ) 
$ \\ 
$z$ & $y$ & $3/2\, ( 3\,{d}^{2}+2 )  ( 4+{d}^{2} ) 
$ \\ 
$z$ & $z$ & $3\, ( 4+{d}^{2} )  ( {d}^{2}+2 )  ( {d}^{2}
+1 ) 
$ \\ 

\hline
\end{tabular}\end{center}
\caption{The quadratic coefficients for the fields up to level 3.  
Off-diagonal terms which are not shown are all 0.}
\label{tab.appendix.quadratic}
\end{table}

\clearpage

{\footnotesize\begin{longtable}{@{}|c@{}c@{}c|p{5.84 in}|}
\hline
$l$ & $m$ & $n$ & $B_{lmn}(d_{1},d_{2},d_{3})$ \\ \hline
\endhead \hline \endfoot
$t$ & $t$ & $t$ & ${\frac {27}{64}}\,\sqrt {3}
$ \\ 
$t$ & $t$ & $x$ & ${\frac {9}{32}}\,d_{{1}}-{\frac {9}{32}}\,d_{{2}}
$ \\ 
$t$ & $t$ & $h$ & $0
$ \\ 
$t$ & $x$ & $x$ & $-1/16\,\sqrt {3} ( -4-d_{{1}}d_{{2}}+{d_{{1}}}^{2}-d_{{3}}d_{{1}}
+d_{{3}}d_{{2}} ) 
$ \\ 
$t$ & $x$ & $h$ & $0
$ \\ 
$t$ & $h$ & $h$ & $1/4\,\sqrt {3}
$ \\ 
$x$ & $x$ & $x$ & $-1/24\, ( d_{{2}}-d_{{3}} )  ( d_{{1}}-d_{{3}}
 )  ( d_{{1}}-d_{{2}} ) 
$ \\ 
$x$ & $x$ & $h$ & $0
$ \\ 
$x$ & $h$ & $h$ & $1/6\,d_{{2}}-1/6\,d_{{3}}
$ \\ 
$h$ & $h$ & $h$ & $0
$ \\ 
$t$ & $t$ & $u$ & ${\frac {11}{64}}\,\sqrt {3}
$ \\ 
$t$ & $t$ & $v$ & $-{\frac {125}{128}}\,\sqrt {3}
$ \\ 
$t$ & $t$ & $w$ & ${\frac {1}{128}}\,\sqrt {3} ( 4\,{d_{{1}}}^{2}-8\,d_{{1}}d_{{2}}+
4\,{d_{{2}}}^{2}+4\,d_{{3}}d_{{1}}+4\,d_{{3}}d_{{2}}-8\,{d_{{3}}}^{2}-
5 ) 
$ \\ 
$t$ & $t$ & $f$ & ${\frac {1}{64}}\,\sqrt {3} ( 2\,d_{{2}}+4\,d_{{3}}{d_{{1}}}^{2}-8
\,d_{{3}}d_{{2}}d_{{1}}+4\,d_{{3}}{d_{{2}}}^{2}-9\,d_{{3}}+2\,d_{{1}}
 ) 
$ \\ 
$t$ & $x$ & $u$ & $-{\frac {11}{96}}\,d_{{1}}+{\frac {11}{96}}\,d_{{3}}
$ \\ 
$t$ & $x$ & $v$ & ${\frac {125}{192}}\,d_{{1}}-{\frac {125}{192}}\,d_{{3}}
$ \\ 
$t$ & $x$ & $w$ & $-1/48\,{d_{{1}}}^{3}-1/6\,d_{{2}}-1/16\,d_{{3}}d_{{2}}d_{{1}}-1/24\,{d
_{{3}}}^{3}+1/48\,{d_{{3}}}^{2}d_{{2}}+1/48\,d_{{3}}{d_{{2}}}^{2}+1/24
\,{d_{{1}}}^{2}d_{{2}}+1/16\,{d_{{3}}}^{2}d_{{1}}+{\frac {37}{192}}\,d
_{{1}}+{\frac {59}{192}}\,d_{{3}}-1/48\,d_{{1}}{d_{{2}}}^{2}
$ \\ 
$t$ & $x$ & $f$ & $-1/24\,d_{{3}}{d_{{1}}}^{3}-1/12\,{d_{{3}}}^{2}d_{{2}}d_{{1}}-1/24\,d_
{{3}}d_{{1}}{d_{{2}}}^{2}-1/48\,d_{{1}}d_{{2}}+1/24\,{d_{{3}}}^{2}{d_{
{2}}}^{2}-{\frac {5}{16}}\,d_{{3}}d_{{2}}+1/24\,{d_{{3}}}^{2}{d_{{1}}}
^{2}-1/48\,{d_{{1}}}^{2}-{\frac {3}{32}}\,{d_{{3}}}^{2}+{\frac {43}{96
}}\,d_{{3}}d_{{1}}+1/12\,d_{{3}}d_{{2}}{d_{{1}}}^{2}+1/3
$ \\ 
$t$ & $h$ & $u$ & $1/6
$ \\ 
$t$ & $h$ & $v$ & $0
$ \\ 
$t$ & $h$ & $w$ & $0
$ \\ 
$t$ & $h$ & $f$ & $0
$ \\ 
$t$ & $u$ & $u$ & ${\frac {19}{576}}\,\sqrt {3}
$ \\ 
$t$ & $u$ & $v$ & $-{\frac {1375}{3456}}\,\sqrt {3}
$ \\ 
$t$ & $u$ & $w$ & ${\frac {11}{3456}}\,\sqrt {3} ( 4\,{d_{{1}}}^{2}-8\,d_{{1}}d_{{2}
}+4\,{d_{{2}}}^{2}+4\,d_{{3}}d_{{1}}+4\,d_{{3}}d_{{2}}-8\,{d_{{3}}}^{2
}-5 ) 
$ \\ 
$t$ & $u$ & $f$ & ${\frac {11}{1728}}\,\sqrt {3} ( 2\,d_{{2}}+4\,d_{{3}}{d_{{1}}}^{2
}-8\,d_{{3}}d_{{2}}d_{{1}}+4\,d_{{3}}{d_{{2}}}^{2}-9\,d_{{3}}+2\,d_{{1
}} ) 
$ \\ 
$t$ & $v$ & $v$ & ${\frac {9475}{2304}}\,\sqrt {3}
$ \\ 
$t$ & $v$ & $w$ & $-{\frac {125}{6912}}\,\sqrt {3} ( 4\,{d_{{1}}}^{2}-8\,d_{{1}}d_{{
2}}+4\,{d_{{2}}}^{2}+4\,d_{{3}}d_{{1}}+4\,d_{{3}}d_{{2}}-8\,{d_{{3}}}^
{2}-5 ) 
$ \\ 
$t$ & $v$ & $f$ & $-{\frac {125}{3456}}\,\sqrt {3} ( 2\,d_{{2}}+4\,d_{{3}}{d_{{1}}}^
{2}-8\,d_{{3}}d_{{2}}d_{{1}}+4\,d_{{3}}{d_{{2}}}^{2}-9\,d_{{3}}+2\,d_{
{1}} ) 
$ \\ 
$t$ & $w$ & $w$ & ${\frac {1}{6912}}\,\sqrt {3} ( 80\,d_{{1}}{d_{{2}}}^{3}+537-16\,{
d_{{3}}}^{3}d_{{2}}+80\,{d_{{3}}}^{3}d_{{1}}-236\,d_{{1}}d_{{2}}-80\,{
d_{{3}}}^{2}d_{{2}}d_{{1}}-48\,{d_{{1}}}^{2}{d_{{2}}}^{2}-48\,{d_{{3}}
}^{2}{d_{{1}}}^{2}-16\,d_{{2}}{d_{{1}}}^{3}+96\,{d_{{3}}}^{2}{d_{{2}}}
^{2}+16\,{d_{{1}}}^{4}-16\,d_{{3}}{d_{{2}}}^{3}+532\,{d_{{3}}}^{2}-296
\,{d_{{1}}}^{2}+532\,{d_{{2}}}^{2}+112\,d_{{3}}d_{{2}}{d_{{1}}}^{2}-32
\,{d_{{2}}}^{4}-32\,{d_{{3}}}^{4}-236\,d_{{3}}d_{{1}}-3368\,d_{{3}}d_{
{2}}-16\,d_{{3}}{d_{{1}}}^{3}-80\,d_{{3}}d_{{1}}{d_{{2}}}^{2} ) 
$ \\ 
$t$ & $w$ & $f$ & ${\frac {1}{3456}}\,\sqrt {3} ( -266\,d_{{1}}+813\,d_{{3}}-16\,{d_
{{2}}}^{3}+16\,{d_{{3}}}^{3}{d_{{2}}}^{2}+16\,{d_{{3}}}^{3}{d_{{1}}}^{
2}+16\,{d_{{3}}}^{2}{d_{{2}}}^{3}-32\,{d_{{3}}}^{2}{d_{{1}}}^{3}-32\,d
_{{3}}{d_{{2}}}^{4}+16\,d_{{3}}{d_{{1}}}^{4}-284\,{d_{{3}}}^{2}d_{{2}}
+572\,d_{{3}}{d_{{2}}}^{2}+8\,{d_{{1}}}^{3}-1546\,d_{{2}}-260\,d_{{3}}
d_{{2}}d_{{1}}-36\,{d_{{3}}}^{3}+16\,{d_{{1}}}^{2}d_{{2}}+336\,{d_{{3}
}}^{2}d_{{1}}-8\,d_{{1}}{d_{{2}}}^{2}-328\,d_{{3}}{d_{{1}}}^{2}-64\,{d
_{{3}}}^{2}d_{{1}}{d_{{2}}}^{2}+80\,{d_{{3}}}^{2}d_{{2}}{d_{{1}}}^{2}-
16\,d_{{3}}d_{{2}}{d_{{1}}}^{3}+80\,d_{{3}}{d_{{2}}}^{3}d_{{1}}-48\,d_
{{3}}{d_{{2}}}^{2}{d_{{1}}}^{2}-32\,{d_{{3}}}^{3}d_{{2}}d_{{1}}
 ) 
$ \\ 
$t$ & $f$ & $f$ & ${\frac {1}{1728}}\,\sqrt {3} ( -768+1109\,d_{{3}}d_{{2}}-18\,{d_{
{2}}}^{2}+320\,d_{{3}}d_{{1}}{d_{{2}}}^{2}-360\,d_{{3}}d_{{2}}{d_{{1}}
}^{2}+4\,{d_{{1}}}^{2}-18\,{d_{{3}}}^{2}+8\,d_{{2}}{d_{{1}}}^{3}-270\,
d_{{3}}d_{{1}}+8\,{d_{{3}}}^{2}{d_{{1}}}^{2}+8\,d_{{3}}{d_{{1}}}^{3}-
240\,{d_{{3}}}^{2}{d_{{2}}}^{2}+16\,{d_{{3}}}^{3}{d_{{2}}}^{3}-270\,d_
{{1}}d_{{2}}+8\,{d_{{1}}}^{2}{d_{{2}}}^{2}-36\,{d_{{3}}}^{3}d_{{2}}+16
\,d_{{3}}{d_{{2}}}^{3}{d_{{1}}}^{2}-32\,d_{{3}}{d_{{2}}}^{2}{d_{{1}}}^
{3}-32\,{d_{{3}}}^{2}d_{{2}}{d_{{1}}}^{3}-32\,{d_{{3}}}^{2}{d_{{2}}}^{
3}d_{{1}}+16\,d_{{3}}d_{{2}}{d_{{1}}}^{4}+16\,{d_{{3}}}^{3}d_{{2}}{d_{
{1}}}^{2}-32\,{d_{{3}}}^{3}{d_{{2}}}^{2}d_{{1}}+64\,{d_{{1}}}^{2}{d_{{
2}}}^{2}{d_{{3}}}^{2}+320\,{d_{{3}}}^{2}d_{{2}}d_{{1}}-36\,d_{{3}}{d_{
{2}}}^{3} ) 
$ \\ 
$x$ & $x$ & $u$ & $-{\frac {11}{432}}\,\sqrt {3} ( -4+d_{{1}}d_{{2}}+{d_{{3}}}^{2}-d
_{{3}}d_{{1}}-d_{{3}}d_{{2}} ) 
$ \\ 
$x$ & $x$ & $v$ & ${\frac {125}{864}}\,\sqrt {3} ( -4+d_{{1}}d_{{2}}+{d_{{3}}}^{2}-d
_{{3}}d_{{1}}-d_{{3}}d_{{2}} ) 
$ \\ 
$x$ & $x$ & $w$ & $-{\frac {1}{864}}\,\sqrt {3} ( 4\,d_{{1}}{d_{{2}}}^{3}-108+12\,{d
_{{3}}}^{3}d_{{2}}+12\,{d_{{3}}}^{3}d_{{1}}-37\,d_{{1}}d_{{2}}-24\,{d_
{{3}}}^{2}d_{{2}}d_{{1}}-8\,{d_{{1}}}^{2}{d_{{2}}}^{2}+4\,d_{{2}}{d_{{
1}}}^{3}-4\,d_{{3}}{d_{{2}}}^{3}+155\,{d_{{3}}}^{2}+16\,{d_{{1}}}^{2}+
16\,{d_{{2}}}^{2}+8\,d_{{3}}d_{{2}}{d_{{1}}}^{2}-8\,{d_{{3}}}^{4}-75\,
d_{{3}}d_{{1}}-75\,d_{{3}}d_{{2}}-4\,d_{{3}}{d_{{1}}}^{3}+8\,d_{{3}}d_
{{1}}{d_{{2}}}^{2} ) 
$ \\ 
$x$ & $x$ & $f$ & $-{\frac {1}{432}}\,\sqrt {3} ( 4\,{d_{{3}}}^{3}{d_{{1}}}^{2}-40\,
d_{{2}}+14\,d_{{3}}{d_{{1}}}^{2}-45\,d_{{3}}d_{{2}}d_{{1}}-4\,{d_{{3}}
}^{2}{d_{{2}}}^{3}-8\,{d_{{3}}}^{3}d_{{2}}d_{{1}}+14\,d_{{3}}{d_{{2}}}
^{2}+4\,d_{{3}}d_{{2}}{d_{{1}}}^{3}-28\,d_{{3}}-4\,{d_{{3}}}^{2}{d_{{1
}}}^{3}+2\,{d_{{1}}}^{2}d_{{2}}+11\,{d_{{3}}}^{2}d_{{2}}+11\,{d_{{3}}}
^{2}d_{{1}}+4\,d_{{3}}{d_{{2}}}^{3}d_{{1}}+4\,{d_{{3}}}^{2}d_{{1}}{d_{
{2}}}^{2}+4\,{d_{{3}}}^{2}d_{{2}}{d_{{1}}}^{2}-8\,d_{{3}}{d_{{2}}}^{2}
{d_{{1}}}^{2}-40\,d_{{1}}+2\,d_{{1}}{d_{{2}}}^{2}+4\,{d_{{3}}}^{3}{d_{
{2}}}^{2}-9\,{d_{{3}}}^{3} ) 
$ \\ 
$x$ & $h$ & $u$ & $1/27\,\sqrt {3} ( d_{{2}}-d_{{3}} ) 
$ \\ 
$x$ & $h$ & $v$ & $0
$ \\ 
$x$ & $h$ & $w$ & $0
$ \\ 
$x$ & $h$ & $f$ & $0
$ \\ 
$h$ & $h$ & $u$ & $1/4\,\sqrt {3}
$ \\ 
$h$ & $h$ & $v$ & $-{\frac {125}{216}}\,\sqrt {3}
$ \\ 
$h$ & $h$ & $w$ & ${\frac {1}{216}}\,\sqrt {3} ( 4\,{d_{{1}}}^{2}-8\,d_{{1}}d_{{2}}+
4\,{d_{{2}}}^{2}+4\,d_{{3}}d_{{1}}+4\,d_{{3}}d_{{2}}-8\,{d_{{3}}}^{2}-
5 ) 
$ \\ 
$h$ & $h$ & $f$ & ${\frac {1}{108}}\,\sqrt {3} ( 2\,d_{{2}}+4\,d_{{3}}{d_{{1}}}^{2}-
8\,d_{{3}}d_{{2}}d_{{1}}+4\,d_{{3}}{d_{{2}}}^{2}-9\,d_{{3}}+2\,d_{{1}}
 ) 
$ \\ 
$x$ & $u$ & $u$ & ${\frac {19}{864}}\,d_{{2}}-{\frac {19}{864}}\,d_{{3}}
$ \\ 
$x$ & $u$ & $v$ & $-{\frac {1375}{5184}}\,d_{{2}}+{\frac {1375}{5184}}\,d_{{3}}
$ \\ 
$x$ & $u$ & $w$ & $-{\frac {11}{432}}\,{d_{{3}}}^{2}d_{{2}}+{\frac {11}{1296}}\,{d_{{1}}}
^{2}d_{{2}}+{\frac {11}{1296}}\,{d_{{2}}}^{3}+{\frac {11}{162}}\,d_{{1
}}-{\frac {11}{1296}}\,{d_{{3}}}^{2}d_{{1}}-{\frac {11}{648}}\,d_{{1}}
{d_{{2}}}^{2}+{\frac {11}{432}}\,d_{{3}}d_{{2}}d_{{1}}-{\frac {11}{
1296}}\,d_{{3}}{d_{{1}}}^{2}-{\frac {407}{5184}}\,d_{{2}}-{\frac {649}
{5184}}\,d_{{3}}+{\frac {11}{648}}\,{d_{{3}}}^{3}
$ \\ 
$x$ & $u$ & $f$ & ${\frac {11}{648}}\,d_{{3}}d_{{2}}{d_{{1}}}^{2}+{\frac {11}{1296}}\,{d_
{{2}}}^{2}+{\frac {11}{324}}\,{d_{{3}}}^{2}d_{{2}}d_{{1}}+{\frac {55}{
432}}\,d_{{3}}d_{{1}}+{\frac {11}{1296}}\,d_{{1}}d_{{2}}+{\frac {11}{
648}}\,d_{{3}}{d_{{2}}}^{3}-{\frac {11}{324}}\,d_{{3}}d_{{1}}{d_{{2}}}
^{2}-{\frac {473}{2592}}\,d_{{3}}d_{{2}}-{\frac {11}{648}}\,{d_{{3}}}^
{2}{d_{{1}}}^{2}-{\frac {11}{648}}\,{d_{{3}}}^{2}{d_{{2}}}^{2}+{\frac 
{11}{288}}\,{d_{{3}}}^{2}-{\frac {11}{81}}
$ \\ 
$x$ & $v$ & $v$ & ${\frac {9475}{3456}}\,d_{{2}}-{\frac {9475}{3456}}\,d_{{3}}
$ \\ 
$x$ & $v$ & $w$ & $-{\frac {125}{324}}\,d_{{1}}+{\frac {4625}{10368}}\,d_{{2}}+{\frac {
7375}{10368}}\,d_{{3}}-{\frac {125}{864}}\,d_{{3}}d_{{2}}d_{{1}}+{
\frac {125}{2592}}\,{d_{{3}}}^{2}d_{{1}}+{\frac {125}{864}}\,{d_{{3}}}
^{2}d_{{2}}+{\frac {125}{2592}}\,d_{{3}}{d_{{1}}}^{2}+{\frac {125}{
1296}}\,d_{{1}}{d_{{2}}}^{2}-{\frac {125}{2592}}\,{d_{{1}}}^{2}d_{{2}}
-{\frac {125}{1296}}\,{d_{{3}}}^{3}-{\frac {125}{2592}}\,{d_{{2}}}^{3}
$ \\ 
$x$ & $v$ & $f$ & $-{\frac {125}{2592}}\,{d_{{2}}}^{2}-{\frac {125}{576}}\,{d_{{3}}}^{2}+
{\frac {125}{648}}\,d_{{3}}d_{{1}}{d_{{2}}}^{2}-{\frac {625}{864}}\,d_
{{3}}d_{{1}}+{\frac {5375}{5184}}\,d_{{3}}d_{{2}}+{\frac {125}{1296}}
\,{d_{{3}}}^{2}{d_{{1}}}^{2}-{\frac {125}{2592}}\,d_{{1}}d_{{2}}+{
\frac {125}{1296}}\,{d_{{3}}}^{2}{d_{{2}}}^{2}-{\frac {125}{1296}}\,d_
{{3}}d_{{2}}{d_{{1}}}^{2}-{\frac {125}{648}}\,{d_{{3}}}^{2}d_{{2}}d_{{
1}}+{\frac {125}{162}}-{\frac {125}{1296}}\,d_{{3}}{d_{{2}}}^{3}
$ \\ 
$x$ & $w$ & $w$ & ${\frac {1}{10368}}\, ( d_{{2}}-d_{{3}} )  ( -32\,{d_{{
2}}}^{4}+80\,d_{{1}}{d_{{2}}}^{3}-16\,d_{{3}}{d_{{2}}}^{3}-80\,d_{{3}}
d_{{1}}{d_{{2}}}^{2}-48\,{d_{{1}}}^{2}{d_{{2}}}^{2}+96\,{d_{{3}}}^{2}{
d_{{2}}}^{2}+1044\,{d_{{2}}}^{2}-16\,d_{{2}}{d_{{1}}}^{3}+112\,d_{{3}}
d_{{2}}{d_{{1}}}^{2}-2088\,d_{{3}}d_{{2}}-16\,{d_{{3}}}^{3}d_{{2}}-
1260\,d_{{1}}d_{{2}}-80\,{d_{{3}}}^{2}d_{{2}}d_{{1}}+1044\,{d_{{3}}}^{
2}+80\,{d_{{3}}}^{3}d_{{1}}+16\,{d_{{1}}}^{4}-2695-48\,{d_{{3}}}^{2}{d
_{{1}}}^{2}+216\,{d_{{1}}}^{2}-32\,{d_{{3}}}^{4}-16\,d_{{3}}{d_{{1}}}^
{3}-1260\,d_{{3}}d_{{1}} ) 
$ \\ 
$x$ & $w$ & $f$ & ${\frac {37}{162}}-{\frac {1}{324}}\,{d_{{3}}}^{4}{d_{{2}}}^{2}+{\frac 
{1}{108}}\,{d_{{3}}}^{2}{d_{{2}}}^{4}-{\frac {1}{324}}\,{d_{{3}}}^{2}{
d_{{1}}}^{4}-{\frac {1}{162}}\,d_{{3}}{d_{{2}}}^{5}-{\frac {1}{324}}\,
{d_{{3}}}^{4}{d_{{1}}}^{2}+{\frac {1}{162}}\,{d_{{3}}}^{3}{d_{{1}}}^{3
}-{\frac {1}{108}}\,d_{{3}}{d_{{2}}}^{3}{d_{{1}}}^{2}-{\frac {1}{324}}
\,d_{{3}}{d_{{2}}}^{2}{d_{{1}}}^{3}-{\frac {1}{324}}\,{d_{{3}}}^{2}d_{
{2}}{d_{{1}}}^{3}-1/36\,{d_{{3}}}^{2}{d_{{2}}}^{3}d_{{1}}+{\frac {5}{
324}}\,d_{{3}}d_{{1}}{d_{{2}}}^{4}+{\frac {1}{324}}\,d_{{3}}d_{{2}}{d_
{{1}}}^{4}-{\frac {1}{81}}\,{d_{{3}}}^{3}d_{{2}}{d_{{1}}}^{2}+{\frac {
1}{162}}\,{d_{{3}}}^{3}{d_{{2}}}^{2}d_{{1}}+{\frac {2}{81}}\,{d_{{1}}}
^{2}{d_{{2}}}^{2}{d_{{3}}}^{2}+{\frac {1}{162}}\,{d_{{3}}}^{4}d_{{2}}d
_{{1}}+{\frac {119}{864}}\,d_{{3}}d_{{1}}-{\frac {55}{864}}\,d_{{1}}d_
{{2}}+{\frac {275}{1296}}\,d_{{3}}{d_{{2}}}^{3}+{\frac {25}{648}}\,{d_
{{3}}}^{2}{d_{{1}}}^{2}-{\frac {107}{648}}\,{d_{{3}}}^{2}{d_{{2}}}^{2}
-{\frac {1}{324}}\,{d_{{2}}}^{4}-1/27\,{d_{{1}}}^{2}+{\frac {7}{216}}
\,d_{{3}}d_{{2}}{d_{{1}}}^{2}+{\frac {181}{1296}}\,{d_{{3}}}^{2}d_{{2}
}d_{{1}}-{\frac {319}{1296}}\,d_{{3}}d_{{1}}{d_{{2}}}^{2}-{\frac {1}{
648}}\,d_{{1}}{d_{{2}}}^{3}+{\frac {1}{324}}\,{d_{{1}}}^{2}{d_{{2}}}^{
2}+{\frac {1}{648}}\,d_{{2}}{d_{{1}}}^{3}+{\frac {5}{216}}\,{d_{{3}}}^
{3}d_{{2}}-{\frac {13}{324}}\,{d_{{3}}}^{3}d_{{1}}-{\frac {1}{648}}\,d
_{{3}}{d_{{1}}}^{3}+{\frac {1}{144}}\,{d_{{3}}}^{4}-{\frac {205}{5184}
}\,{d_{{3}}}^{2}-{\frac {581}{2592}}\,{d_{{2}}}^{2}-{\frac {1193}{5184
}}\,d_{{3}}d_{{2}}
$ \\ 
$x$ & $f$ & $f$ & ${\frac {1}{2592}}\, ( d_{{2}}-d_{{3}} )  ( 608+16\,{d_
{{3}}}^{3}{d_{{2}}}^{3}+16\,d_{{3}}{d_{{2}}}^{3}{d_{{1}}}^{2}-32\,d_{{
3}}{d_{{2}}}^{2}{d_{{1}}}^{3}-32\,{d_{{3}}}^{2}d_{{2}}{d_{{1}}}^{3}-32
\,{d_{{3}}}^{2}{d_{{2}}}^{3}d_{{1}}+16\,d_{{3}}d_{{2}}{d_{{1}}}^{4}+16
\,{d_{{3}}}^{3}d_{{2}}{d_{{1}}}^{2}-32\,{d_{{3}}}^{3}{d_{{2}}}^{2}d_{{
1}}+64\,{d_{{1}}}^{2}{d_{{2}}}^{2}{d_{{3}}}^{2}-334\,d_{{3}}d_{{1}}-
334\,d_{{1}}d_{{2}}-36\,d_{{3}}{d_{{2}}}^{3}+8\,{d_{{3}}}^{2}{d_{{1}}}
^{2}-112\,{d_{{3}}}^{2}{d_{{2}}}^{2}-188\,{d_{{1}}}^{2}-232\,d_{{3}}d_
{{2}}{d_{{1}}}^{2}+192\,{d_{{3}}}^{2}d_{{2}}d_{{1}}+192\,d_{{3}}d_{{1}
}{d_{{2}}}^{2}+8\,{d_{{1}}}^{2}{d_{{2}}}^{2}+8\,d_{{2}}{d_{{1}}}^{3}-
36\,{d_{{3}}}^{3}d_{{2}}+8\,d_{{3}}{d_{{1}}}^{3}-18\,{d_{{3}}}^{2}-18
\,{d_{{2}}}^{2}+565\,d_{{3}}d_{{2}} ) 
$ \\ 
$h$ & $u$ & $u$ & $0
$ \\ 
$h$ & $u$ & $v$ & $-{\frac {125}{324}}
$ \\ 
$h$ & $u$ & $w$ & ${\frac {1}{81}}\,{d_{{1}}}^{2}-{\frac {2}{81}}\,d_{{1}}d_{{2}}+{\frac 
{1}{81}}\,{d_{{2}}}^{2}+{\frac {1}{81}}\,d_{{3}}d_{{1}}+{\frac {1}{81}
}\,d_{{3}}d_{{2}}-{\frac {2}{81}}\,{d_{{3}}}^{2}-{\frac {5}{324}}
$ \\ 
$h$ & $u$ & $f$ & ${\frac {1}{81}}\,d_{{1}}+{\frac {1}{81}}\,d_{{2}}-1/18\,d_{{3}}+{
\frac {2}{81}}\,d_{{3}}{d_{{1}}}^{2}-{\frac {4}{81}}\,d_{{3}}d_{{2}}d_
{{1}}+{\frac {2}{81}}\,d_{{3}}{d_{{2}}}^{2}
$ \\ 
$h$ & $v$ & $v$ & $0
$ \\ 
$h$ & $v$ & $w$ & $0
$ \\ 
$h$ & $v$ & $f$ & $0
$ \\ 
$h$ & $w$ & $w$ & $0
$ \\ 
$h$ & $w$ & $f$ & $0
$ \\ 
$h$ & $f$ & $f$ & $0
$ \\ 
$u$ & $u$ & $u$ & ${\frac {1}{192}}\,\sqrt {3}
$ \\ 
$u$ & $u$ & $v$ & $-{\frac {2375}{31104}}\,\sqrt {3}
$ \\ 
$u$ & $u$ & $w$ & ${\frac {19}{31104}}\,\sqrt {3} ( 4\,{d_{{1}}}^{2}-8\,d_{{1}}d_{{2
}}+4\,{d_{{2}}}^{2}+4\,d_{{3}}d_{{1}}+4\,d_{{3}}d_{{2}}-8\,{d_{{3}}}^{
2}-5 ) 
$ \\ 
$u$ & $u$ & $f$ & ${\frac {19}{15552}}\,\sqrt {3} ( 2\,d_{{2}}+4\,d_{{3}}{d_{{1}}}^{
2}-8\,d_{{3}}d_{{2}}d_{{1}}+4\,d_{{3}}{d_{{2}}}^{2}-9\,d_{{3}}+2\,d_{{
1}} ) 
$ \\ 
$u$ & $v$ & $v$ & ${\frac {104225}{62208}}\,\sqrt {3}
$ \\ 
$u$ & $v$ & $w$ & $-{\frac {1375}{186624}}\,\sqrt {3} ( 4\,{d_{{1}}}^{2}-8\,d_{{1}}d
_{{2}}+4\,{d_{{2}}}^{2}+4\,d_{{3}}d_{{1}}+4\,d_{{3}}d_{{2}}-8\,{d_{{3}
}}^{2}-5 ) 
$ \\ 
$u$ & $v$ & $f$ & $-{\frac {1375}{93312}}\,\sqrt {3} ( 2\,d_{{2}}+4\,d_{{3}}{d_{{1}}
}^{2}-8\,d_{{3}}d_{{2}}d_{{1}}+4\,d_{{3}}{d_{{2}}}^{2}-9\,d_{{3}}+2\,d
_{{1}} ) 
$ \\ 
$u$ & $w$ & $w$ & ${\frac {11}{186624}}\,\sqrt {3} ( 80\,d_{{1}}{d_{{2}}}^{3}+537-16
\,{d_{{3}}}^{3}d_{{2}}+80\,{d_{{3}}}^{3}d_{{1}}-236\,d_{{1}}d_{{2}}-80
\,{d_{{3}}}^{2}d_{{2}}d_{{1}}-48\,{d_{{1}}}^{2}{d_{{2}}}^{2}-48\,{d_{{
3}}}^{2}{d_{{1}}}^{2}-16\,d_{{2}}{d_{{1}}}^{3}+96\,{d_{{3}}}^{2}{d_{{2
}}}^{2}+16\,{d_{{1}}}^{4}-16\,d_{{3}}{d_{{2}}}^{3}+532\,{d_{{3}}}^{2}-
296\,{d_{{1}}}^{2}+532\,{d_{{2}}}^{2}+112\,d_{{3}}d_{{2}}{d_{{1}}}^{2}
-32\,{d_{{2}}}^{4}-32\,{d_{{3}}}^{4}-236\,d_{{3}}d_{{1}}-3368\,d_{{3}}
d_{{2}}-16\,d_{{3}}{d_{{1}}}^{3}-80\,d_{{3}}d_{{1}}{d_{{2}}}^{2}
 ) 
$ \\ 
$u$ & $w$ & $f$ & ${\frac {11}{93312}}\,\sqrt {3} ( 16\,{d_{{3}}}^{2}{d_{{2}}}^{3}+8
\,{d_{{1}}}^{3}+16\,{d_{{3}}}^{3}{d_{{2}}}^{2}+16\,d_{{3}}{d_{{1}}}^{4
}-16\,{d_{{2}}}^{3}-36\,{d_{{3}}}^{3}-48\,d_{{3}}{d_{{2}}}^{2}{d_{{1}}
}^{2}-260\,d_{{3}}d_{{2}}d_{{1}}+80\,d_{{3}}{d_{{2}}}^{3}d_{{1}}-16\,d
_{{3}}d_{{2}}{d_{{1}}}^{3}+80\,{d_{{3}}}^{2}d_{{2}}{d_{{1}}}^{2}-32\,{
d_{{3}}}^{3}d_{{2}}d_{{1}}+336\,{d_{{3}}}^{2}d_{{1}}+16\,{d_{{3}}}^{3}
{d_{{1}}}^{2}-284\,{d_{{3}}}^{2}d_{{2}}+16\,{d_{{1}}}^{2}d_{{2}}-328\,
d_{{3}}{d_{{1}}}^{2}+572\,d_{{3}}{d_{{2}}}^{2}-32\,{d_{{3}}}^{2}{d_{{1
}}}^{3}-32\,d_{{3}}{d_{{2}}}^{4}-266\,d_{{1}}-8\,d_{{1}}{d_{{2}}}^{2}-
1546\,d_{{2}}+813\,d_{{3}}-64\,{d_{{3}}}^{2}d_{{1}}{d_{{2}}}^{2}
 ) 
$ \\ 
$u$ & $f$ & $f$ & ${\frac {11}{46656}}\,\sqrt {3} ( -768+16\,{d_{{3}}}^{3}{d_{{2}}}^
{3}-270\,d_{{3}}d_{{1}}-270\,d_{{1}}d_{{2}}-36\,d_{{3}}{d_{{2}}}^{3}+8
\,{d_{{3}}}^{2}{d_{{1}}}^{2}-240\,{d_{{3}}}^{2}{d_{{2}}}^{2}+16\,d_{{3
}}{d_{{2}}}^{3}{d_{{1}}}^{2}-32\,d_{{3}}{d_{{2}}}^{2}{d_{{1}}}^{3}-32
\,{d_{{3}}}^{2}d_{{2}}{d_{{1}}}^{3}-32\,{d_{{3}}}^{2}{d_{{2}}}^{3}d_{{
1}}+16\,d_{{3}}d_{{2}}{d_{{1}}}^{4}+16\,{d_{{3}}}^{3}d_{{2}}{d_{{1}}}^
{2}-32\,{d_{{3}}}^{3}{d_{{2}}}^{2}d_{{1}}+64\,{d_{{1}}}^{2}{d_{{2}}}^{
2}{d_{{3}}}^{2}+4\,{d_{{1}}}^{2}-360\,d_{{3}}d_{{2}}{d_{{1}}}^{2}+320
\,{d_{{3}}}^{2}d_{{2}}d_{{1}}+320\,d_{{3}}d_{{1}}{d_{{2}}}^{2}+8\,{d_{
{1}}}^{2}{d_{{2}}}^{2}+8\,d_{{2}}{d_{{1}}}^{3}-36\,{d_{{3}}}^{3}d_{{2}
}+8\,d_{{3}}{d_{{1}}}^{3}-18\,{d_{{3}}}^{2}-18\,{d_{{2}}}^{2}+1109\,d_
{{3}}d_{{2}} ) 
$ \\ 
$v$ & $v$ & $v$ & $-{\frac {219775}{13824}}\,\sqrt {3}
$ \\ 
$v$ & $v$ & $w$ & ${\frac {9475}{124416}}\,\sqrt {3} ( 4\,{d_{{1}}}^{2}-8\,d_{{1}}d_
{{2}}+4\,{d_{{2}}}^{2}+4\,d_{{3}}d_{{1}}+4\,d_{{3}}d_{{2}}-8\,{d_{{3}}
}^{2}-5 ) 
$ \\ 
$v$ & $v$ & $f$ & ${\frac {9475}{62208}}\,\sqrt {3} ( 2\,d_{{2}}+4\,d_{{3}}{d_{{1}}}
^{2}-8\,d_{{3}}d_{{2}}d_{{1}}+4\,d_{{3}}{d_{{2}}}^{2}-9\,d_{{3}}+2\,d_
{{1}} ) 
$ \\ 
$v$ & $w$ & $w$ & $-{\frac {125}{373248}}\,\sqrt {3} ( 80\,d_{{1}}{d_{{2}}}^{3}+537-
16\,{d_{{3}}}^{3}d_{{2}}+80\,{d_{{3}}}^{3}d_{{1}}-236\,d_{{1}}d_{{2}}-
80\,{d_{{3}}}^{2}d_{{2}}d_{{1}}-48\,{d_{{1}}}^{2}{d_{{2}}}^{2}-48\,{d_
{{3}}}^{2}{d_{{1}}}^{2}-16\,d_{{2}}{d_{{1}}}^{3}+96\,{d_{{3}}}^{2}{d_{
{2}}}^{2}+16\,{d_{{1}}}^{4}-16\,d_{{3}}{d_{{2}}}^{3}+532\,{d_{{3}}}^{2
}-296\,{d_{{1}}}^{2}+532\,{d_{{2}}}^{2}+112\,d_{{3}}d_{{2}}{d_{{1}}}^{
2}-32\,{d_{{2}}}^{4}-32\,{d_{{3}}}^{4}-236\,d_{{3}}d_{{1}}-3368\,d_{{3
}}d_{{2}}-16\,d_{{3}}{d_{{1}}}^{3}-80\,d_{{3}}d_{{1}}{d_{{2}}}^{2}
 ) 
$ \\ 
$v$ & $w$ & $f$ & $-{\frac {125}{186624}}\,\sqrt {3} ( 16\,{d_{{3}}}^{2}{d_{{2}}}^{3
}+8\,{d_{{1}}}^{3}+16\,{d_{{3}}}^{3}{d_{{2}}}^{2}+16\,d_{{3}}{d_{{1}}}
^{4}-16\,{d_{{2}}}^{3}-36\,{d_{{3}}}^{3}-48\,d_{{3}}{d_{{2}}}^{2}{d_{{
1}}}^{2}-260\,d_{{3}}d_{{2}}d_{{1}}+80\,d_{{3}}{d_{{2}}}^{3}d_{{1}}-16
\,d_{{3}}d_{{2}}{d_{{1}}}^{3}+80\,{d_{{3}}}^{2}d_{{2}}{d_{{1}}}^{2}-32
\,{d_{{3}}}^{3}d_{{2}}d_{{1}}+336\,{d_{{3}}}^{2}d_{{1}}+16\,{d_{{3}}}^
{3}{d_{{1}}}^{2}-284\,{d_{{3}}}^{2}d_{{2}}+16\,{d_{{1}}}^{2}d_{{2}}-
328\,d_{{3}}{d_{{1}}}^{2}+572\,d_{{3}}{d_{{2}}}^{2}-32\,{d_{{3}}}^{2}{
d_{{1}}}^{3}-32\,d_{{3}}{d_{{2}}}^{4}-266\,d_{{1}}-8\,d_{{1}}{d_{{2}}}
^{2}-1546\,d_{{2}}+813\,d_{{3}}-64\,{d_{{3}}}^{2}d_{{1}}{d_{{2}}}^{2}
 ) 
$ \\ 
$v$ & $f$ & $f$ & $-{\frac {125}{93312}}\,\sqrt {3} ( -768+16\,{d_{{3}}}^{3}{d_{{2}}
}^{3}-270\,d_{{3}}d_{{1}}-270\,d_{{1}}d_{{2}}-36\,d_{{3}}{d_{{2}}}^{3}
+8\,{d_{{3}}}^{2}{d_{{1}}}^{2}-240\,{d_{{3}}}^{2}{d_{{2}}}^{2}+16\,d_{
{3}}{d_{{2}}}^{3}{d_{{1}}}^{2}-32\,d_{{3}}{d_{{2}}}^{2}{d_{{1}}}^{3}-
32\,{d_{{3}}}^{2}d_{{2}}{d_{{1}}}^{3}-32\,{d_{{3}}}^{2}{d_{{2}}}^{3}d_
{{1}}+16\,d_{{3}}d_{{2}}{d_{{1}}}^{4}+16\,{d_{{3}}}^{3}d_{{2}}{d_{{1}}
}^{2}-32\,{d_{{3}}}^{3}{d_{{2}}}^{2}d_{{1}}+64\,{d_{{1}}}^{2}{d_{{2}}}
^{2}{d_{{3}}}^{2}+4\,{d_{{1}}}^{2}-360\,d_{{3}}d_{{2}}{d_{{1}}}^{2}+
320\,{d_{{3}}}^{2}d_{{2}}d_{{1}}+320\,d_{{3}}d_{{1}}{d_{{2}}}^{2}+8\,{
d_{{1}}}^{2}{d_{{2}}}^{2}+8\,d_{{2}}{d_{{1}}}^{3}-36\,{d_{{3}}}^{3}d_{
{2}}+8\,d_{{3}}{d_{{1}}}^{3}-18\,{d_{{3}}}^{2}-18\,{d_{{2}}}^{2}+1109
\,d_{{3}}d_{{2}} ) 
$ \\ 
$w$ & $w$ & $w$ & $-{\frac {1}{373248}}\,\sqrt {3} ( -24963-6384\,{d_{{1}}}^{4}+
15600\,d_{{1}}{d_{{2}}}^{3}-192\,{d_{{3}}}^{5}d_{{2}}-192\,d_{{2}}{d_{
{1}}}^{5}-384\,{d_{{3}}}^{4}{d_{{1}}}^{2}+896\,{d_{{2}}}^{3}{d_{{1}}}^
{3}+896\,{d_{{3}}}^{3}{d_{{1}}}^{3}+896\,{d_{{3}}}^{3}{d_{{2}}}^{3}-
27392\,d_{{3}}d_{{1}}-27392\,d_{{1}}d_{{2}}+15600\,d_{{3}}{d_{{2}}}^{3
}-18432\,{d_{{3}}}^{2}{d_{{1}}}^{2}-18432\,{d_{{3}}}^{2}{d_{{2}}}^{2}+
1152\,d_{{3}}d_{{1}}{d_{{2}}}^{4}+1152\,{d_{{3}}}^{4}d_{{2}}d_{{1}}-
960\,d_{{3}}{d_{{2}}}^{3}{d_{{1}}}^{2}-960\,d_{{3}}{d_{{2}}}^{2}{d_{{1
}}}^{3}-960\,{d_{{3}}}^{2}d_{{2}}{d_{{1}}}^{3}-960\,{d_{{3}}}^{2}{d_{{
2}}}^{3}d_{{1}}+1152\,d_{{3}}d_{{2}}{d_{{1}}}^{4}-960\,{d_{{3}}}^{3}d_
{{2}}{d_{{1}}}^{2}-960\,{d_{{3}}}^{3}{d_{{2}}}^{2}d_{{1}}+2688\,{d_{{1
}}}^{2}{d_{{2}}}^{2}{d_{{3}}}^{2}+44800\,{d_{{1}}}^{2}-6384\,d_{{3}}d_
{{2}}{d_{{1}}}^{2}-6384\,{d_{{3}}}^{2}d_{{2}}d_{{1}}-6384\,d_{{3}}d_{{
1}}{d_{{2}}}^{2}+15600\,{d_{{3}}}^{3}d_{{1}}-384\,{d_{{1}}}^{4}{d_{{2}
}}^{2}-384\,{d_{{3}}}^{4}{d_{{2}}}^{2}-192\,d_{{3}}{d_{{1}}}^{5}-192\,
{d_{{3}}}^{5}d_{{1}}-384\,{d_{{1}}}^{2}{d_{{2}}}^{4}-384\,{d_{{3}}}^{2
}{d_{{2}}}^{4}-192\,d_{{1}}{d_{{2}}}^{5}-384\,{d_{{3}}}^{2}{d_{{1}}}^{
4}-192\,d_{{3}}{d_{{2}}}^{5}-18432\,{d_{{1}}}^{2}{d_{{2}}}^{2}+15600\,
d_{{2}}{d_{{1}}}^{3}+15600\,{d_{{3}}}^{3}d_{{2}}+15600\,d_{{3}}{d_{{1}
}}^{3}-6384\,{d_{{2}}}^{4}+128\,{d_{{2}}}^{6}+128\,{d_{{1}}}^{6}+44800
\,{d_{{3}}}^{2}-6384\,{d_{{3}}}^{4}+128\,{d_{{3}}}^{6}+44800\,{d_{{2}}
}^{2}-27392\,d_{{3}}d_{{2}} ) 
$ \\ 
$w$ & $w$ & $f$ & $-{\frac {1}{186624}}\,\sqrt {3} ( 9952\,{d_{{3}}}^{2}{d_{{2}}}^{3
}-616\,{d_{{3}}}^{3}+3032\,{d_{{2}}}^{3}-3312\,{d_{{3}}}^{3}{d_{{1}}}^
{2}-15868\,{d_{{3}}}^{2}d_{{1}}+37160\,d_{{3}}{d_{{1}}}^{2}+1576\,d_{{
1}}{d_{{2}}}^{2}+1576\,{d_{{1}}}^{2}d_{{2}}+9952\,{d_{{3}}}^{2}{d_{{1}
}}^{3}-160\,{d_{{1}}}^{2}{d_{{2}}}^{3}+96\,d_{{1}}{d_{{2}}}^{4}+96\,{d
_{{1}}}^{4}d_{{2}}-160\,{d_{{2}}}^{2}{d_{{1}}}^{3}-176\,{d_{{3}}}^{4}d
_{{1}}+64\,{d_{{3}}}^{4}{d_{{1}}}^{3}+1280\,{d_{{3}}}^{3}{d_{{2}}}^{2}
{d_{{1}}}^{2}+128\,{d_{{3}}}^{5}d_{{2}}d_{{1}}-64\,{d_{{3}}}^{4}d_{{2}
}{d_{{1}}}^{2}-832\,{d_{{3}}}^{3}{d_{{2}}}^{3}d_{{1}}-640\,{d_{{3}}}^{
2}{d_{{2}}}^{2}{d_{{1}}}^{3}-384\,d_{{3}}{d_{{2}}}^{4}{d_{{1}}}^{2}-
640\,{d_{{3}}}^{2}{d_{{2}}}^{3}{d_{{1}}}^{2}-832\,{d_{{3}}}^{3}d_{{2}}
{d_{{1}}}^{3}+64\,{d_{{1}}}^{5}-1842\,d_{{1}}-176\,{d_{{3}}}^{4}d_{{2}
}-3312\,{d_{{3}}}^{3}{d_{{2}}}^{2}+3032\,{d_{{1}}}^{3}-15868\,{d_{{3}}
}^{2}d_{{2}}+192\,{d_{{3}}}^{3}{d_{{2}}}^{4}+64\,{d_{{3}}}^{4}{d_{{2}}
}^{3}+192\,{d_{{3}}}^{3}{d_{{1}}}^{4}-6672\,d_{{3}}{d_{{2}}}^{4}-320\,
{d_{{3}}}^{2}{d_{{2}}}^{5}-64\,{d_{{3}}}^{5}{d_{{2}}}^{2}-6672\,d_{{3}
}{d_{{1}}}^{4}-320\,{d_{{3}}}^{2}{d_{{1}}}^{5}-64\,{d_{{3}}}^{5}{d_{{1
}}}^{2}+128\,d_{{3}}{d_{{2}}}^{6}+128\,d_{{3}}{d_{{1}}}^{6}-9984\,{d_{
{3}}}^{2}d_{{1}}{d_{{2}}}^{2}-64\,{d_{{3}}}^{4}{d_{{2}}}^{2}d_{{1}}-
384\,d_{{3}}{d_{{2}}}^{2}{d_{{1}}}^{4}+960\,{d_{{3}}}^{2}d_{{1}}{d_{{2
}}}^{4}+37160\,d_{{3}}{d_{{2}}}^{2}+64\,{d_{{2}}}^{5}-9984\,{d_{{3}}}^
{2}d_{{2}}{d_{{1}}}^{2}+15536\,d_{{3}}d_{{2}}{d_{{1}}}^{3}+15536\,d_{{
3}}{d_{{2}}}^{3}d_{{1}}-192\,d_{{3}}d_{{2}}{d_{{1}}}^{5}-192\,d_{{3}}d
_{{1}}{d_{{2}}}^{5}-33776\,d_{{3}}d_{{2}}d_{{1}}-17728\,d_{{3}}{d_{{2}
}}^{2}{d_{{1}}}^{2}+6896\,{d_{{3}}}^{3}d_{{2}}d_{{1}}+896\,d_{{3}}{d_{
{2}}}^{3}{d_{{1}}}^{3}+960\,{d_{{3}}}^{2}{d_{{1}}}^{4}d_{{2}}+144\,{d_
{{3}}}^{5}-1842\,d_{{2}}-3871\,d_{{3}} ) 
$ \\ 
$w$ & $f$ & $f$ & $-{\frac {1}{93312}}\,\sqrt {3} ( 4352-9050\,{d_{{2}}}^{2}-1862\,d
_{{3}}d_{{1}}-2456\,d_{{3}}{d_{{2}}}^{3}+1472\,{d_{{3}}}^{2}{d_{{1}}}^
{2}+2984\,d_{{3}}{d_{{1}}}^{3}+1664\,d_{{1}}{d_{{2}}}^{3}+32\,{d_{{1}}
}^{4}{d_{{2}}}^{2}-352\,{d_{{3}}}^{4}{d_{{2}}}^{2}-32\,{d_{{1}}}^{2}{d
_{{2}}}^{4}-352\,{d_{{3}}}^{2}{d_{{2}}}^{4}+144\,{d_{{3}}}^{5}d_{{2}}+
64\,d_{{2}}{d_{{1}}}^{5}+144\,d_{{3}}{d_{{2}}}^{5}+2984\,d_{{2}}{d_{{1
}}}^{3}-64\,d_{{3}}{d_{{2}}}^{5}{d_{{1}}}^{2}-22276\,d_{{3}}d_{{1}}{d_
{{2}}}^{2}+29744\,d_{{3}}d_{{2}}{d_{{1}}}^{2}-3056\,d_{{3}}{d_{{2}}}^{
3}{d_{{1}}}^{2}+9984\,d_{{3}}{d_{{2}}}^{2}{d_{{1}}}^{3}+9984\,{d_{{3}}
}^{2}d_{{2}}{d_{{1}}}^{3}+3248\,{d_{{3}}}^{2}{d_{{2}}}^{3}d_{{1}}-112
\,d_{{3}}d_{{1}}{d_{{2}}}^{4}+14249\,d_{{3}}d_{{2}}+7304\,{d_{{3}}}^{2
}{d_{{2}}}^{2}+1664\,{d_{{3}}}^{3}d_{{1}}-9050\,{d_{{3}}}^{2}+20\,{d_{
{1}}}^{2}+128\,{d_{{3}}}^{4}{d_{{2}}}^{4}-2456\,{d_{{3}}}^{3}d_{{2}}+
64\,d_{{3}}{d_{{1}}}^{5}-32\,{d_{{3}}}^{4}{d_{{1}}}^{2}-64\,{d_{{3}}}^
{3}{d_{{1}}}^{3}-1862\,d_{{1}}d_{{2}}-64\,{d_{{2}}}^{3}{d_{{1}}}^{3}+
496\,{d_{{3}}}^{3}{d_{{2}}}^{3}+1472\,{d_{{1}}}^{2}{d_{{2}}}^{2}+72\,{
d_{{3}}}^{4}-6960\,d_{{3}}d_{{2}}{d_{{1}}}^{4}-3056\,{d_{{3}}}^{3}d_{{
2}}{d_{{1}}}^{2}+3248\,{d_{{3}}}^{3}{d_{{2}}}^{2}d_{{1}}-13248\,{d_{{1
}}}^{2}{d_{{2}}}^{2}{d_{{3}}}^{2}-22276\,{d_{{3}}}^{2}d_{{2}}d_{{1}}-
112\,{d_{{3}}}^{4}d_{{2}}d_{{1}}-64\,{d_{{3}}}^{5}d_{{2}}{d_{{1}}}^{2}
+896\,{d_{{3}}}^{2}{d_{{1}}}^{4}{d_{{2}}}^{2}+128\,{d_{{3}}}^{2}{d_{{2
}}}^{5}d_{{1}}+192\,{d_{{3}}}^{3}d_{{2}}{d_{{1}}}^{4}+72\,{d_{{2}}}^{4
}-64\,{d_{{3}}}^{3}{d_{{2}}}^{5}+32\,{d_{{1}}}^{4}-704\,{d_{{3}}}^{2}{
d_{{2}}}^{3}{d_{{1}}}^{3}+128\,d_{{3}}d_{{2}}{d_{{1}}}^{6}+32\,{d_{{3}
}}^{2}{d_{{1}}}^{4}-64\,{d_{{3}}}^{5}{d_{{2}}}^{3}-320\,d_{{3}}{d_{{2}
}}^{2}{d_{{1}}}^{5}+128\,{d_{{3}}}^{5}{d_{{2}}}^{2}d_{{1}}+192\,d_{{3}
}{d_{{2}}}^{3}{d_{{1}}}^{4}-320\,{d_{{3}}}^{2}d_{{2}}{d_{{1}}}^{5}+64
\,{d_{{3}}}^{4}d_{{2}}{d_{{1}}}^{3}+768\,{d_{{3}}}^{3}{d_{{2}}}^{3}{d_
{{1}}}^{2}-704\,{d_{{3}}}^{3}{d_{{2}}}^{2}{d_{{1}}}^{3}-192\,{d_{{3}}}
^{4}{d_{{2}}}^{3}d_{{1}}-192\,{d_{{3}}}^{3}{d_{{2}}}^{4}d_{{1}}+64\,d_
{{3}}{d_{{2}}}^{4}{d_{{1}}}^{3} ) 
$ \\ 
$f$ & $f$ & $f$ & ${\frac {1}{46656}}\,\sqrt {3} ( -1576\,{d_{{3}}}^{2}{d_{{2}}}^{3}
-36\,{d_{{3}}}^{3}-36\,{d_{{2}}}^{3}-1576\,{d_{{3}}}^{3}{d_{{1}}}^{2}+
9094\,{d_{{3}}}^{2}d_{{1}}+9094\,d_{{3}}{d_{{1}}}^{2}+9094\,d_{{1}}{d_
{{2}}}^{2}+9094\,{d_{{1}}}^{2}d_{{2}}-1576\,{d_{{3}}}^{2}{d_{{1}}}^{3}
-1576\,{d_{{1}}}^{2}{d_{{2}}}^{3}-72\,d_{{1}}{d_{{2}}}^{4}-72\,{d_{{1}
}}^{4}d_{{2}}-1576\,{d_{{2}}}^{2}{d_{{1}}}^{3}-72\,{d_{{3}}}^{4}d_{{1}
}+32\,{d_{{3}}}^{4}{d_{{1}}}^{3}+192\,{d_{{3}}}^{3}{d_{{2}}}^{2}{d_{{1
}}}^{2}-144\,{d_{{3}}}^{5}d_{{2}}d_{{1}}+288\,{d_{{3}}}^{4}d_{{2}}{d_{
{1}}}^{2}-688\,{d_{{3}}}^{3}{d_{{2}}}^{3}d_{{1}}+192\,{d_{{3}}}^{2}{d_
{{2}}}^{2}{d_{{1}}}^{3}+288\,d_{{3}}{d_{{2}}}^{4}{d_{{1}}}^{2}+192\,{d
_{{3}}}^{2}{d_{{2}}}^{3}{d_{{1}}}^{2}-688\,{d_{{3}}}^{3}d_{{2}}{d_{{1}
}}^{3}-4352\,d_{{1}}-72\,{d_{{3}}}^{4}d_{{2}}-1576\,{d_{{3}}}^{3}{d_{{
2}}}^{2}+128\,{d_{{3}}}^{4}{d_{{2}}}^{3}{d_{{1}}}^{2}+128\,{d_{{3}}}^{
3}{d_{{2}}}^{4}{d_{{1}}}^{2}-36\,{d_{{1}}}^{3}+128\,{d_{{3}}}^{2}{d_{{
2}}}^{3}{d_{{1}}}^{4}+64\,{d_{{3}}}^{3}d_{{2}}{d_{{1}}}^{5}+64\,d_{{3}
}{d_{{2}}}^{3}{d_{{1}}}^{5}+128\,{d_{{3}}}^{2}{d_{{2}}}^{4}{d_{{1}}}^{
3}+128\,{d_{{3}}}^{4}{d_{{2}}}^{2}{d_{{1}}}^{3}+64\,d_{{3}}{d_{{2}}}^{
5}{d_{{1}}}^{3}-128\,{d_{{3}}}^{2}{d_{{2}}}^{5}{d_{{1}}}^{2}-128\,{d_{
{3}}}^{4}d_{{2}}{d_{{1}}}^{4}+64\,{d_{{3}}}^{3}{d_{{2}}}^{5}d_{{1}}-
128\,d_{{3}}{d_{{1}}}^{4}{d_{{2}}}^{4}-128\,{d_{{3}}}^{4}{d_{{2}}}^{4}
d_{{1}}-128\,{d_{{3}}}^{2}{d_{{2}}}^{2}{d_{{1}}}^{5}+9094\,{d_{{3}}}^{
2}d_{{2}}+32\,{d_{{3}}}^{3}{d_{{2}}}^{4}+32\,{d_{{3}}}^{4}{d_{{2}}}^{3
}+32\,{d_{{3}}}^{3}{d_{{1}}}^{4}-72\,d_{{3}}{d_{{2}}}^{4}-72\,d_{{3}}{
d_{{1}}}^{4}-968\,{d_{{3}}}^{2}d_{{1}}{d_{{2}}}^{2}+288\,{d_{{3}}}^{4}
{d_{{2}}}^{2}d_{{1}}+288\,d_{{3}}{d_{{2}}}^{2}{d_{{1}}}^{4}+288\,{d_{{
3}}}^{2}d_{{1}}{d_{{2}}}^{4}+9094\,d_{{3}}{d_{{2}}}^{2}+32\,{d_{{2}}}^
{3}{d_{{1}}}^{4}+64\,{d_{{3}}}^{5}{d_{{2}}}^{3}d_{{1}}+64\,{d_{{3}}}^{
5}d_{{2}}{d_{{1}}}^{3}-128\,{d_{{3}}}^{5}{d_{{2}}}^{2}{d_{{1}}}^{2}+
128\,{d_{{3}}}^{3}{d_{{2}}}^{2}{d_{{1}}}^{4}-384\,{d_{{3}}}^{3}{d_{{2}
}}^{3}{d_{{1}}}^{3}+32\,{d_{{2}}}^{4}{d_{{1}}}^{3}-968\,{d_{{3}}}^{2}d
_{{2}}{d_{{1}}}^{2}+4264\,d_{{3}}d_{{2}}{d_{{1}}}^{3}+4264\,d_{{3}}{d_
{{2}}}^{3}d_{{1}}-144\,d_{{3}}d_{{2}}{d_{{1}}}^{5}-144\,d_{{3}}d_{{1}}
{d_{{2}}}^{5}-29493\,d_{{3}}d_{{2}}d_{{1}}-968\,d_{{3}}{d_{{2}}}^{2}{d
_{{1}}}^{2}+4264\,{d_{{3}}}^{3}d_{{2}}d_{{1}}-688\,d_{{3}}{d_{{2}}}^{3
}{d_{{1}}}^{3}+288\,{d_{{3}}}^{2}{d_{{1}}}^{4}d_{{2}}-4352\,d_{{2}}-
4352\,d_{{3}} ) 
$ \\ 
\end{longtable}}
\label{tab.appendix.cubic}


\subsection*{Acknowledgment:}
This work was completed with support from the Natural Sciences and Engineering
Council of Canada.

\bibliographystyle{JHEP}
\bibliography{refs}

\end{document}